\documentclass[a4paper]{jpconf}
\usepackage{graphicx}
\usepackage{amssymb}
\begin{document}
\title{Three puzzles from nuclear astrophysics}

\author{W. C. Haxton}

\address{ Department of Physics, University of California, Berkeley, 
 and Lawrence Berkeley National Laboratory, 
 MC-7300, Berkeley, CA 94720 }

\ead{haxton@berkeley.edu}

\begin{abstract}
I discuss three open problems in astrophysics where nuclear physics can make important
contributions:  the solar abundance problem, dark matter particle detection, and the origin 
of the r-process elements.
\end{abstract}

It is a great pleasure to take part in this celebration of Jerry Draayer's career in physics and
to describe three problems in nuclear astrophysics that I find of interest.
The descriptions here will be brief, as detailed recent
references are available.  The topics are 1) the solar abundance problem and its
implications for the solar model and future solar neutrino experiments; 2) the elastic 
scattering of dark matter (DM) particles off nuclei;
and 3) the possibility that an early, neutrino-driven r-process
(rapid-neutron-capture process) could, in
combination with neutron star mergers, account for what we know about the galactic
evolution of r-process metals.

\section{The solar abundance problem}
The Standard Solar Model (SSM) \cite{Book89,HRS13} assumes a homogeneous zero-age Sun because of the 
assumption that the proto-Sun passed through a fully convective
Hayashi phase during gas-cloud collapse.   The initial composition is divided into 
hydrogen X$_\mathrm{ini}$, helium Y$_\mathrm{ini}$, and metals Z$_\mathrm{ini}$.
The relative abundances of
the metals (elements heavier than helium) can be taken from
meteoritic data (refractory elements) and from analyses of photospheric absorption lines
(volatiles such as C, N, O, Ne, Ar).  As X$_\mathrm{ini}$+Y$_\mathrm{ini}$+Z$_\mathrm{ini}$=1,
the initial solar composition can be fully specified by adjusting Y$_\mathrm{ini}$,
Z$_\mathrm{ini}$, as well as one other parameter, the mixing length $\alpha_\mathrm{MLT}$,
until the SSM, evolved forward by 4.6 Gyrs, reproduces the present solar radius $R_\odot$, 
luminosity $L_\odot$, and surface helium abundance Y$_S$.
The metals are important contributors to the opacity and consequently influence a
variety of solar properties.

Traditionally the analysis of photospheric absorption lines was done with a 1D model
that did not explicitly take into account surface stratification, velocities, and inhomogeneities \cite{GS98}.
More recently 3D parameter-free methods have been introduced, improving
the consistency of line analyses \cite{AGSS09}.   This approach led to a significant reduction
in Z$_\mathrm{ini}$, from $\sim$ 0.0169 to $\sim$ 0.0122, a change that makes the Sun
more consistent with similar stars in the local neighborhood.  However the agreement between the SSM and
various helioseismic properties, such as the sound speed profile and the location of the 
convective zone boundary, deteriorated with the reduced Z$_\mathrm{ini}$
(see Fig. 1).  The high-energy $^8$B neutrino flux is also affected,
with the high-metallicity GS98-SFII SSM flux being $\sim$ 22\% higher than 
low-metallicity AGSS09-SFII SSM flux \cite{HRS13}.

\begin{figure}
\begin{center}
\includegraphics[width=10cm]{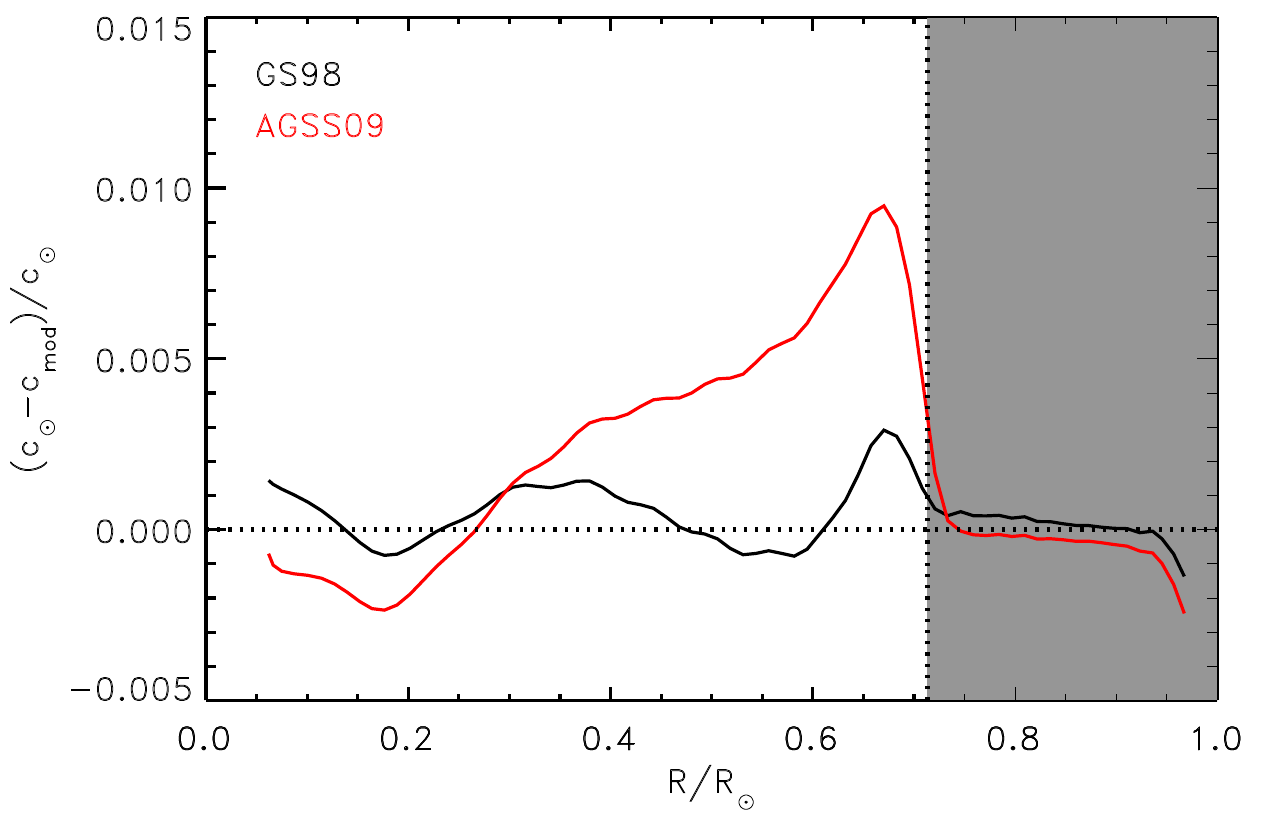}
\caption{The differences between the sound-speed profile determined from helioseismology and
those predicted by the
high-metallicity GS98-SFII and low-metallicity AGSS09-SFII SSMs.}
\label{fig1:c}
\end{center}
\end{figure}

The solar abundance problem can be summarized as follows:  There is a significant discrepancy
between a SSM tuned to our best description of the solar surface (a 3D treatment
of photoabsorption lines) and one tuned to reproduce
properties of the solar interior (helioseismic observables).  The two
models have very different Z$_\mathrm{ini}$s.

This problem is potentially related to two others illustrated in Fig. 2.  The first is the anomalous 
composition \cite{Nordlund09} of the
gaseous giant planets, Jupiter, Saturn, Uranus, and Neptune.   Jupiter and Saturn are enriched
in C and N by factors of 4-7, relative to the Sun.  The process of planetary formation, which
is thought to have occurred late in the evolution of the solar system when only $\sim$ 5\%
of the nebular gas remained in a thin disk, appears to have scoured approximately 40-90 
earth masses (M$_\oplus$) of metal from the disk \cite{Guillot05}.  The second comes from comparing the Sun's 
composition with that of solar twins \cite{Melendez09}, where systematic
differences in metal abundances track condensation temperatures, with the solar ratio of volatiles to
refractories being 0.05 to 0.10 dex higher than the average twin ratio.  

\begin{figure}
\begin{center}
\includegraphics[width=16cm]{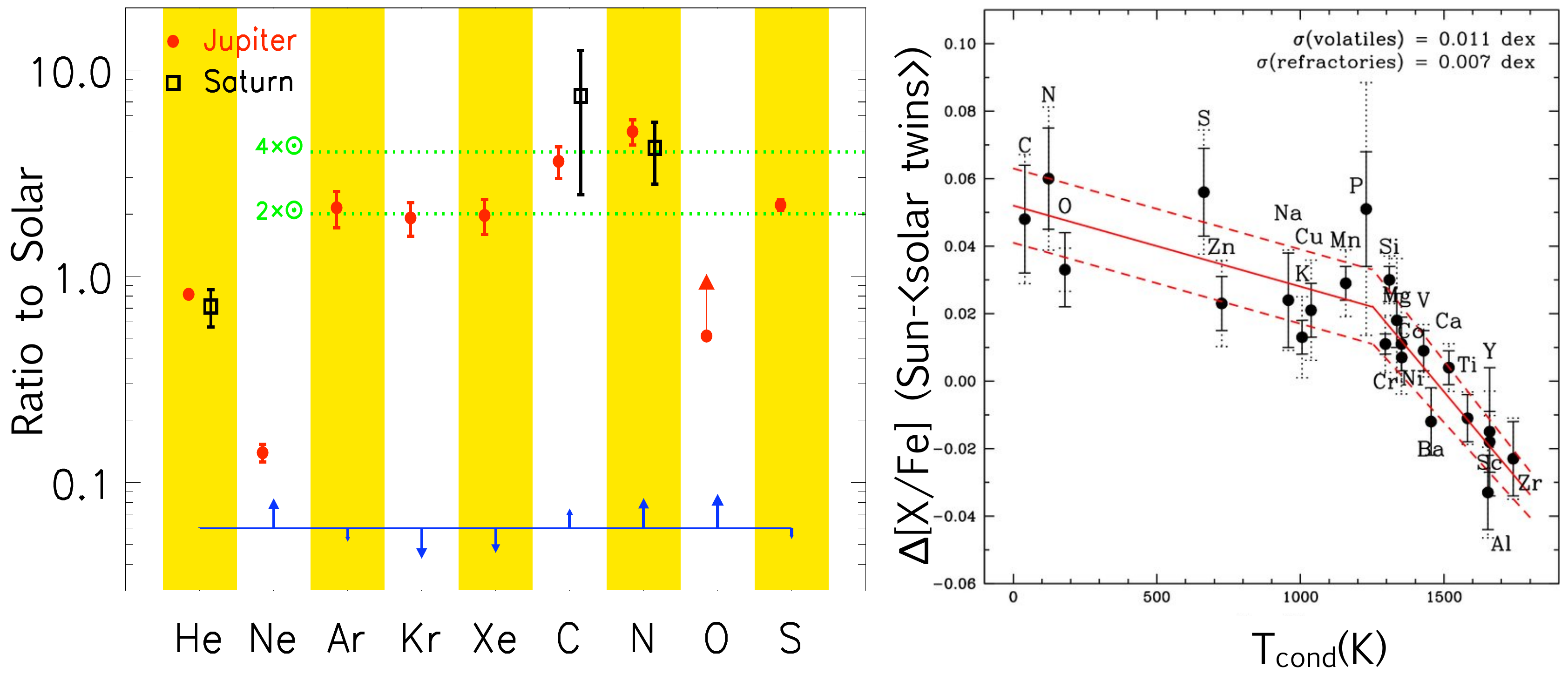}
\caption{Left panel: Data from the Galileo and Cassini missions showing abundances in the gaseous
atmospheres of Jupiter and Saturn, relative to solar. From \cite{Guillot05}.   Right panel: Deviation between solar metallicities,
normalized to iron, from corresponding average quantities for solar twins, displayed as
a function of the condensation temperature of the metal.  From \cite{Melendez09}.}
\label{fig2:twins}
\end{center}
\end{figure}

These three problems could have a common origin, planetary disk processes that concentrate ice, dust, and
other metal-rich material in the planets.  If the depleted gas from which the metals were scrubbed were deposited onto the
Sun after the Sun had developed its radiative core (so that the convective envelope at the time resembled
the modern one, which contains less than 3\% of the Sun's mass), then significant dilution of the
convective zone could have occurred.  This could produce a Sun with a higher metallicity
radiative core and lower metallicity convective surface, invalidating the SSM's assumption of
zero-age homogeneity.  (Indeed, the excess metal in the planets is similar to the net difference
in GS98-SFII and AGSS09-SFII SSM convective-zone metal, about 40 M$_\oplus$.) 

There has been a significant amount of recent work focused on this conjecture.  The nuclear physics connections include:
\begin{enumerate}
\item Neutrino flux measurements that are sensitive to changes in solar core temperature $T_c$ of 1\%, and
thus to metallicity differences comparable to those being debated.
\item The development of accreting nonstandard solar models (NSSMs), and the demonstration that such models are sharply constrained by solar neutrino and helioseismic measurements.
\item The possibility that future solar neutrino flux measurements could provide a direct
measurement of the C+N content of the solar core.
\end{enumerate}
In the case of the neutrino fluxes, the most sensitivity to $T_c$ is exhibited by the $^8$B neutrinos.
However, despite current measurements that determine this flux to $\sim$ 3\% and despite
SSM model uncertainties of $\sim$ 14\% in this flux that exceed the nominal discrepancy of nearly 22\% between
the competing models \cite{HRS13}, little can be concluded:  the AGSS09-SFII and GS98-SFII SSM
global fits to the solar neutrino data have identical $\chi^2$s.  A metallicity intermediate between
the two models would optimize the fit.

The impact of accretion on both neutrino fluxes and helioseismic variables has been evaluated in
NSSMs.  The work to date \cite{SHP11} varied the time, mass, and composition of the accreted material,
but has not allowed the composition to vary in time -- a simplification that one could question
in view of the right panel of Fig. \ref{fig2:twins}.  The results are summarized in Figs. 10 and 11
of \cite{HRS13}.  Helioseismic and neutrino observables tightly constrain such NSSMs.  
The envisioned candidate solution to the solar abundance problem -- a low-Z surface consistent
with AGSS09 abundances and a high-Z interior consistent with GS98 abundances -- can be achieved with 
metal-free or metal-poor accretion and modest accreted masses M$_\mathrm{ac} \sim 0.01$M$_\odot$.
Such models can bring the surface helium abundance into agreement with the data and produce
some improvement in the sound-speed figure-of-merit $\langle \delta c/c \rangle$ and in 
the comparison of model and measured neutrino fluxes.  However the lower surface metallicity
that accompanies such models forces the convective-zone boundary $R_S$ to move outward, 
contradicting results from helioseismology.

The third point above --  checking interior solar abundances by using solar neutrinos as a probe --
is possible because of the CN neutrinos.
While the Sun produces 99\% of its energy through the pp chain, proton burning also
takes place through the CN I cycle, which produces significant fluxes of $^{13}$N ($\sim 2.9 \times
10^8$/cm$^2$/s) and $^{15}$O ($\sim 2.2 \times 10^8$/cm$^2$/s) neutrinos.   These fluxes,
like the $^8$B neutrino flux, are very sensitive to $T_c$, which can be
changed by varying any of the approximately 19 SSM input parameters, according to their
assigned uncertainties.  These uncertainties include the abundances of key metals.
But in addition to the response in $T_c$ due to changes in metallicity, the CN cycle has
an additional linear dependence on the C+N abundance because these elements
catalyze the proton burning.

In \cite {HS08} a strategy for using the neutrino flux to measure the core C+N content was
developed in which a weighted ratio of $^{15}$O and $^8$B fluxes is formed that is nearly
independent of $T_c$, isolating this linear dependence on C+N.  The result (including
recent improvements in the analysis) is
\begin{equation}
{\phi(^{15}\mathrm{O} \over \phi(^{15}\mathrm{O})^\mathrm{SSM}} = \left[ {\phi(^{8}\mathrm{B} \over \phi(^{8}\mathrm{B})^\mathrm{SSM}} \right]^{0.729} x_{\mathrm{C+N}} \left[1 \pm 0.006(\mathrm{SSM})
\pm 0.027(\mathrm{D}) \pm 0.099(\mathrm{nucl}) \pm 0.032 (\theta_{12}) \right].
\label{eq:HS}
\end{equation}
The first factor on the right --
effectively our $T_c$ ``thermometer" -- is now known experimentally to 2\%.  The uncertainty denoted ``SSM"  
reflects the effects on this relationship when 18 of 19 SSM input parameters are varied.
The response of 0.6\% reflects the success in constructing a $T_c$-independent relationship.  
But there is sensitivity to the 19th parameter, the diffusion coefficient D:
Eq. (\ref{eq:HS}) relates a contemporary neutrino flux measurement to the primordial
C+N abundance.  The composition of the modern Sun's core is altered by the gravitational settling of C+N over
4.6 Gyr, an effect that the SSM predicts is uncertain to 2.7\%.  The dominant uncertainty in Eq. (\ref{eq:HS}) of
9.9\% comes from nuclear physics, particularly the S-factors for $^7$Be(p,$\gamma$) and
$^{14}$N(p,$\gamma$).  These S-factor uncertainties can be reduced \cite{SFII}: 
with a factor-of-two improvement, nuclear physics would no longer dominate the error analysis.  Finally there is a
3.2\% uncertainty due to weak interaction parameters.

On the left appears a flux that is currently not measured, but could be determined to a precision
of about 7\% in the upcoming solar neutrino experiment SNO+ \cite{Chen}, a larger and significantly deeper
version of Borexino.  The depth -- the experiment is under construction in the former SNO
cavity at SNOLab -- is important, potentially reducing backgrounds due to 
cosmogenic $^{11}$C by a factor of $\sim$ 70, relative to Borexino.  When all errors are combined
in quadrature, one finds that the solar core abundance of C+N could be determined to $\sim$ 13\%,
which can be compared to the current debate over differences in Z$_\mathrm{ini}$ of $\gtrsim$ 30\%.

SNO+ could allow, for the first time in stellar astrophysics, a direct experimental comparison
between core and surface compositions, thereby testing the SSM assumption of initial
homogeneity.  If it were shown that planetary formation affects host-star surface metallicity 
in a characteristic way -- e.g., producing an elevated ratio of volatiles to refractories -- this
could be important in identifying stars that are
more likely to host planets.

\section{The nuclear physics of dark matter detection}
Several independent astrophysical observations suggest that 85\% of gravitating matter is
dark, residing beyond the Standard Model.  This DM must be long-lived or stable,
and must lack strong couplings to itself and to baryons.  A leading candidate for the DM is
weakly interacting massive particles (WIMPs), slow-moving massive particles (e.g.,
10 GeV to 1 TeV) that can elastically scattering off nuclear targets, transferring 
significant momentum ($\sim$ 100 MeV).  Such DM particles can be detected through
nuclear recoil in carefully designed detectors mounted in low-background environments,
deep underground \cite{Feng10}.

Indeed, there is a great deal of experimental activity in DM detection, including several claims of possible
discovery, and disagreements among experimentalists about the compatibility of the claims
with existing limits.  Various natural targets are being used,
containing isotopes such as $^{19}$F, $^{23}$Na, $^{70,72,73,74,76}$Ge,
$^{127}$I, and $^{128,129,130,131,132,134,136}$Xe.  Thus targets include isotopes sensitive to vector
(J$\geq$1/2) and tensor (J$\geq$1) interactions, such as $^{19}$F(1/2$^+$) and $^{129}$Xe(1/2$^+$);
and $^{23}$Na(3/2$^+$), $^{73}$Ge(9/2$^+$), $^{127}$I(5/2$^+$), and $^{131}$Xe(3/2$^+$).

DM particles could in principle scatter off any scalar, vector, or tensor static nuclear moment.
However analyses almost always treat the nuclear response in the point-nucleus limit,
despite momentum transfers comparable to the inverse nuclear size, so that the scattering is
\begin{eqnarray}
\mathrm{spin}-\mathrm{independent~(S.I.)} &\Rightarrow& \langle g.s. | \sum_{i=1}^A (a_0^F + a_1^F \tau_3(i)) | g.s. \rangle~~~\mathrm{or} \nonumber \\
\mathrm{spin}-\mathrm{dependent~(S.D.)} &\Rightarrow&  \langle g.s. | \sum_{i=1}^A \vec{\sigma}(i) (a_0^{GT} + a_1^{GT} \tau_3(i)) | g.s. \rangle.
\end{eqnarray}
Most often theoretical treatments are ``top-down" -- calculations motivated by a specific high-energy theory, 
for which a nuclear response involving a S.I. or S.D. response is then derived.

Recently a different approach was taken in which the most general Galilean-invariant
effective DM-nucleon interaction was derived, including operators up to quadratic order in momenta and
exchanges of particles of spin-1 or less \cite{Liam12}.  That interaction has 11 couplings
($\times$ 2 if isospin is included), in contrast to the  S.I. and S.D. cases discussed above:
\begin{eqnarray}
L_{ET} &=& a_1 1 + a_2 \vec{v}^\perp \cdot \vec{v}^\perp + a_3 \vec{S}_N \cdot (\vec{q} \times \vec{v}^\perp)
+a_4 \vec{S}_\chi \cdot \vec{S}_N +i a_5 \vec{S}_\chi \cdot (\vec{q} \times \vec{v}^\perp) 
+a_6 \vec{S}_\chi \cdot \vec{q}~ \vec{S}_N \cdot \vec{q} \nonumber \\
&+&a_7 \vec{S}_N \cdot \vec{v}^\perp+a_8 \vec{S}_\chi \cdot \vec{v}^\perp +i a_9 \vec{S}_\chi \cdot (\vec{S}_N \times \vec{q})
+i a_{10} \vec{S}_N \cdot \vec{q} + i a_{11} \vec{S_\chi} \cdot \vec{q}
\label{eq:ET}
\end{eqnarray}
Here $\vec{S}_N$ and $\vec{S}_\chi$ are the nucleon and DM particle spin, $\vec{q}$ is the three-momentum
transfer, and $\vec{v}^\perp$ is the Galilean-invariant velocity operator defined in \cite{Liam12}.  The coefficients
of the eleven effective operators include an isospin dependence that has been suppressed,
e.g., $a_1 \equiv a_1^0 + a_1^1 \tau_3(i)$, to allow for arbitrary couplings to protons and neutrons.

This starting point provides a vehicle for determining the corresponding most general elastic
nuclear response, using the prescription that A-body nuclear currents and charges are the 
sums over the corresponding one-body nucleon currents and charges.  One finds
\begin{eqnarray}
H_{ET} &=& \sum_{i=1}^A  \left[  l_0(i) \delta(\vec{x}-\vec{x}_i) +  l_0^A(i){1 \over 2M} \left( -{1 \over i} \overleftarrow{\nabla}_i
\cdot \vec{\sigma}(i) \delta(\vec{x}-\vec{x}_i)+ \delta(\vec{x}-\vec{x}_i) \vec{\sigma}(i) \cdot {1 \over i}
\overrightarrow{\nabla}_i \right) \right. \nonumber \\
&+&  \vec{l}_5(i) \cdot \vec{\sigma}(i) \delta(\vec{x}-\vec{x}_i) +\vec{l}_M(i) \cdot {1 \over 2M} \left(
-{1 \over i} \overleftarrow{\nabla}_i \delta(\vec{x}-\vec{x}_i) + \delta (\vec{x}-\vec{x}_i) {1 \over i}
\overrightarrow{\nabla}_i \right) \nonumber \\
&+&  \left. \vec{l}_E(i) \cdot {1 \over 2M} \left( \overleftarrow{\nabla}_i \times
\vec{\sigma}(i) \delta(\vec{x}-\vec{x}_i) + \delta(\vec{x}-\vec{x}_i) \vec{\sigma}(i) \times
\overrightarrow{\nabla}_i \right)  \right]
\label{eq:H}
\end{eqnarray}
Here each of the associated WIMP amplitudes is a function of the various $a_i$ appearing in Eq. (\ref{eq:ET})
(including the isospin dependence), e.g.,  
\begin{eqnarray}
\label{eq:ls}
l_0(i)  &=&  \left( a_1^0  -i (\vec{q} \times \vec{S}_\chi) \cdot \vec{v}_T^\perp ~a_5^0
+ \vec{S}_\chi \cdot \vec{v}_T^\perp ~a_8^0 + i \vec{q} \cdot \vec{S}_\chi ~a_{11}^0 \right) \nonumber \\
&+&  \left( a_1^1  -i (\vec{q} \times \vec{S}_\chi) \cdot \vec{v}_T^\perp ~ a_5^1
+ \vec{S}_\chi \cdot \vec{v}_T^\perp ~ a_8^1 + i \vec{q} \cdot \vec{S}_\chi ~ a_{11}^1 \right) \tau_3(i) \equiv l_0^0+ l_0^1 \tau_3(i).
\end{eqnarray}

Equation (\ref{eq:H}) involves five nuclear densities, each of which is familiar from studies
of semi-leptonic electroweak interactions,
\begin{enumerate}
\item The isoscalar and isovector vector charge operators $\rho_V^0 (\vec{x}) = \sum_{i=1}^A \delta(\vec{x}-\vec{x}_i)$ and 
 $\rho_V^1 (\vec{x}) = \sum_{i=1}^A \delta(\vec{x}-\vec{x}_i) \tau_3(i)$ familiar from elastic electron scattering;
 \item The isoscalar axial-vector charge operator $\rho_A^0 (\vec{x}) = \sum_{i=1}^A {1 \over 2M} \left( -{1 \over i} \overleftarrow{\nabla}_i \cdot \vec{\sigma}(i) \delta(\vec{x}-\vec{x}_i)+\right.$   $\left. \delta(\vec{x}-\vec{x}_i) \vec{\sigma}(i) \cdot {1 \over i} \overrightarrow{\nabla}_i \right)$ and the corresponding isovector operator that mediates axial-charge $0^+ \leftrightarrow 0^-~ \beta$ decay;
 \item The isoscalar axial-vector spin operator $\vec{j}_A^{~0} (\vec{x}) = \sum_{i=1}^A \vec{\sigma}(i) \delta(\vec{x}-\vec{x}_i)$ and its isovector counterpart, familiar from
 inelastic neutrino reactions in the long-wavelength limit (LWL); 
 \item The isoscalar vector velocity current $\vec{j}_{V,v}^{~0} (\vec{x}) = \sum_{i=1}^A {1 \over 2M} \left( -{1 \over i} \overleftarrow{\nabla}_i  \delta(\vec{x}-\vec{x}_i)+ \delta(\vec{x}-\vec{x}_i) {1 \over i} \overrightarrow{\nabla}_i \right)$  
 and its isovector counterpart, familiar from electron scattering; and
 \item The isoscalar vector spin-velocity current $\vec{j}_{V,sv}^{~0} (\vec{x}) = \sum_{i=1}^A {1 \over 2M} \left( \overleftarrow{\nabla}_i \times \vec{\sigma}(i) \delta(\vec{x}-\vec{x}_i)+\right.$   $\left. \delta(\vec{x}-\vec{x}_i) \vec{\sigma}(i) \times\overrightarrow{\nabla}_i \right)$ and its isovector counterpart.
 \end{enumerate}
 The vector spin-velocity current appears in Serot's \cite{Serot} treatment of semi-leptonic
 electroweak interactions to order $1/M^2$.  In that context, because of the 
 properties of this current under time reversal, it 
 is accompanying by a factor of the energy transfer $q_0$ that vanishes for elastic transitions.
 
 Equation (\ref{eq:H}) follows from $L_{ET}$, but could have been written down directly as a
 nuclear-level effective theory, as it includes all possible local one-body operators of rank $\leq 1$ that can be constructed
 with $\vec{\sigma}$, $\tau_3$, and one gradient.
 
\begin{figure}
\begin{center}
\includegraphics[width=16cm]{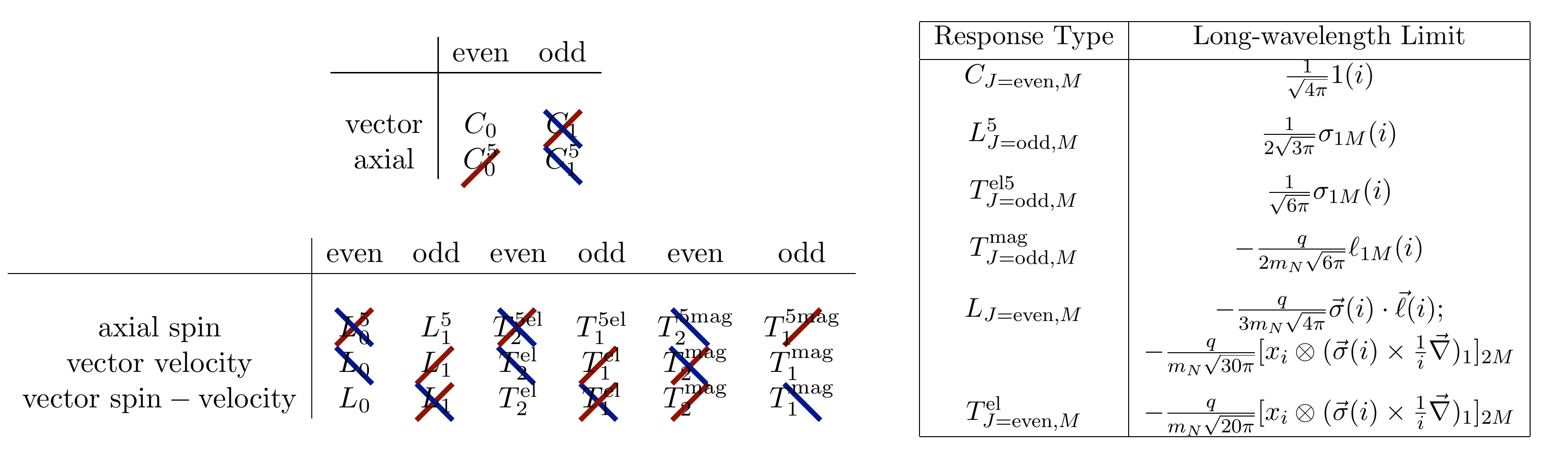}
\caption{Left: Parity and time-reversal constraints on multipoles mediating WIMP elastic 
scattering.  Only the lowest even and lowest odd multipoles are shown, as all even and all odd multipoles
transform similarly.  Multipoles eliminated by parity (time reversal)
are indicated by a red (blue)  strike. 
Right: Table of the surviving parity- and time-reversal-allowed responses, and their LWLs.}
\label{fig3:multipoles}
\end{center}
\end{figure}

 Nuclear ground states are approximate eigenstates of parity and time-reversal.  Thus, 
 ignoring small components of the wave function that
 break these symmetries, one should retain only multipoles generated from the nuclear densities that 
 transform appropriately under parity and time reversal.  The charge operators of Eq. (\ref{eq:H}) generate even and odd
 multipoles:  because all even and all odd multipoles transform similarly, it is sufficient to consider
 only the J=0 and J=1 cases.   The current operators generate longitudinal, transverse electric, and
 transverse magnetic multipoles.  Again, we can limit our consideration to the lowest
 contributing even and odd multipoles, as the generalization to other cases is immediate.  The
 multipoles eliminated by symmetry considerations are indicated in Fig. \ref{fig3:multipoles}.
 
Fig. \ref{fig3:multipoles} also gives the LWLs of the surviving multipoles, which are
 even vector charge operators ($C_0$, C$_2$,...), even vector spin-velocity
longitudinal operators ($L_0$, $L_2$,....), odd axial spin longitudinal operators ($L_1^5$, $L_3^5$, ...),
even vector spin-velocity electric operators ($T_2^\mathrm{el}$, $T_4^\mathrm{el}$, ...),
odd axial spin electric operators ($T_1^{5 \mathrm{el}}$, $T_3^{5 \mathrm{el}}$,...),  and 
odd vector spin magnetic operators ($T_1^{\mathrm{mag}}$, $T_3^{\mathrm{mag}}$,...).  Thus
there are six independent response functions -- not simply the two (S.D., S.I.)  conventionally
treated.  
 
The $C_{J=\mathrm{even,M}}$ operator of Fig. \ref{fig3:multipoles} is the standard S.I. response,
while $L^5_{J=\mathrm{odd},M}$ and $T^{\mathrm{el}5}_{J=\mathrm{odd},M}$ are
the longitudinal and transverse components that are usually summed to give the S.D. response.  
These two contributors to the S.D. response have distinct form factors and couple in distinct ways to WIMP amplitudes, however.
The  transverse magnetic response $T^\mathrm{mag}_{J=\mathrm{odd},M}$ is new, generated by the vector
velocity current.   In the LWL the operator is
proportional to the  nuclear orbital angular momentum operator $\vec{\ell}(i)$, and thus is explicitly 
connected with the composite nature of the nucleon (apparent from the
accompanying factor of $q$ in the right panel of Fig \ref{fig3:multipoles}).
Thus there are three responses -- one proportional to $\vec{\ell}$ and two proportional to $\vec{\sigma}$
in the LWL  -- that
transform under rotations as $\langle J_i | \vec{J}_M | J_i \rangle$.   The fifth and sixth responses
of Fig. \ref{fig3:multipoles} are generated by the spin-velocity current and also reflect nucleon
compositeness.  They transform as longitudinal and transverse electric projections of the vector spin-velocity
current $\sim \delta(\vec{x}-\vec{x}_i) \vec{\sigma}(i) \times \vec{\nabla}$, respectively  -- just as the two S.D. responses
arise as the longitudinal and transverse electric projections of the axial spin current.  The first,
$L_{J=\mathrm{even},M}$,
generates a scalar nuclear response proportional to the spin-orbit interaction $\vec{\sigma}(i) \cdot \vec{\ell}(i)$ as well as a tensor contribution, in the LWL.  The scalar response is present for all nuclei (that is, 
regardless of ground-state spin, like the S.I.
charge coupling).  But its properties are very different from those of the S.I. operator:
$\vec{\sigma}(i) \cdot \vec{\ell}(i)$ produces a coherent isoscalar contribution over closed spin-orbit partner 
shells, e.g., the closed $1f_{7/2}$
shell for Ge isotopes and the closed $1g_{9/2}$ shell for Xe or I.    The sixth response
function, $T^\mathrm{el}_{J=\mathrm{even},M}$, is the transverse electric projection of the spin-velocity current.

\begin{figure}
\begin{center}
\includegraphics[width=11cm]{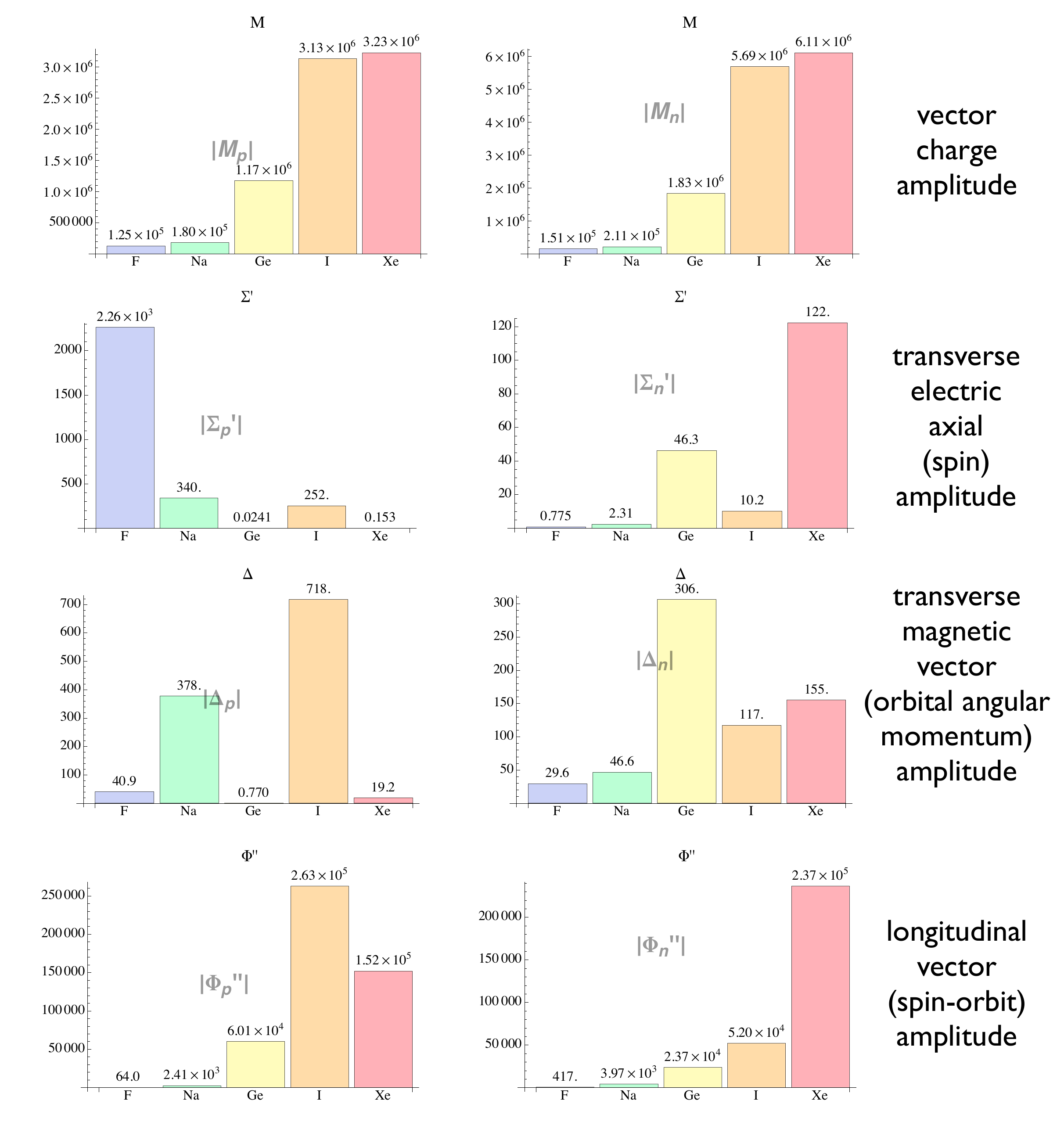}
\caption{Magnitudes of response function amplitudes are compared for four of the response functions discussed
in the text (the S.I. vector response, the S.D. transverse electric axial response, the transverse magnetic
vector response,
and the longitudinal vector response).  The
isospin couplings are to protons only (left panels) and neutrons only (right panels).
From \cite{Liam12}.}
\label{fig4:responses}
\end{center}
\end{figure} 

Figure \ref{fig4:responses} shows response functions calculated in the shell model for natural
targets of F, Na, Ge, I, and Xe \cite{Liam12} -- these responses could be labeled as S.I., S.D. (transverse axial electric), 
orbital angular momentum, and spin-orbit, in the LWL.  Very large differences in
experimental sensitivity appear between detector materials, as the interaction and thus the 
response is varied -- the space of possibilities
is much larger than in the standard S.I./S.D. description.  One might think this is unfortunate -- the 
correlations between different experiments will be harder to anticipate -- but I would argue with this view.
We know nothing about DM interactions, but we know a lot about nuclear responses.  The six nuclear
responses provide six nuclear dials that can be used to probe the nature of DM.
The experimental DM community might benefit from greater interaction with the nuclear physics
community, in formulating strategies for exploiting the available nuclear responses
to determine DM particle properties.

\section{The search for the r-process site}
The search for the site or sites of the r-process, the rapid-neutron-capture process thought to be responsible
for synthesizing about half the heavy elements, has been a long and frustrating one.  For the past
two decades the leading candidate site \cite{Woosley92} has been the neutrino-driven wind that blows off the
proto-neutron star surface in a core-collapse supernova.   As the slightly neutron-rich, high-entropy 
nucleon gas expands off the star and cools, the protons are locked up in an alpha freeze-out.
A small number of seed nuclei can be synthesized through alpha-induced reactions.  The remaining
free neutrons then capture on these seeds to produce heavy nuclei.  If the necessary neutron/seed ratio
of $\gtrsim 100$ can be maintained, the synthesis can continue up to the transuranics.

Unfortunately a number of difficulties have arisen in explorations of the neutrino-wind mechanism.  It is
difficult to keep the number of seeds from growing due to reactions such as 
$\alpha+\alpha + n \rightarrow {}^9\mathrm{Be} + \gamma$ and $3 \alpha \rightarrow {}^{12}\mathrm{C} + \gamma$.
It is also difficult to maintain the needed neutron fluence.
The wind is needed as the ejection mechanism, but every $\nu_e + n \rightarrow p + e^-$ reaction results
in the net loss of two neutrons via $2n+2p \rightarrow {}^4\mathrm{He} + \gamma$.   Consequently, it was recently
concluded that the net production in the standard neutrino wind scenario was limited to some N=50
closed-shell nuclei and to some light p-nuclei \cite{Roberts10}.

\begin{figure}
\begin{center}
\includegraphics[width=16cm]{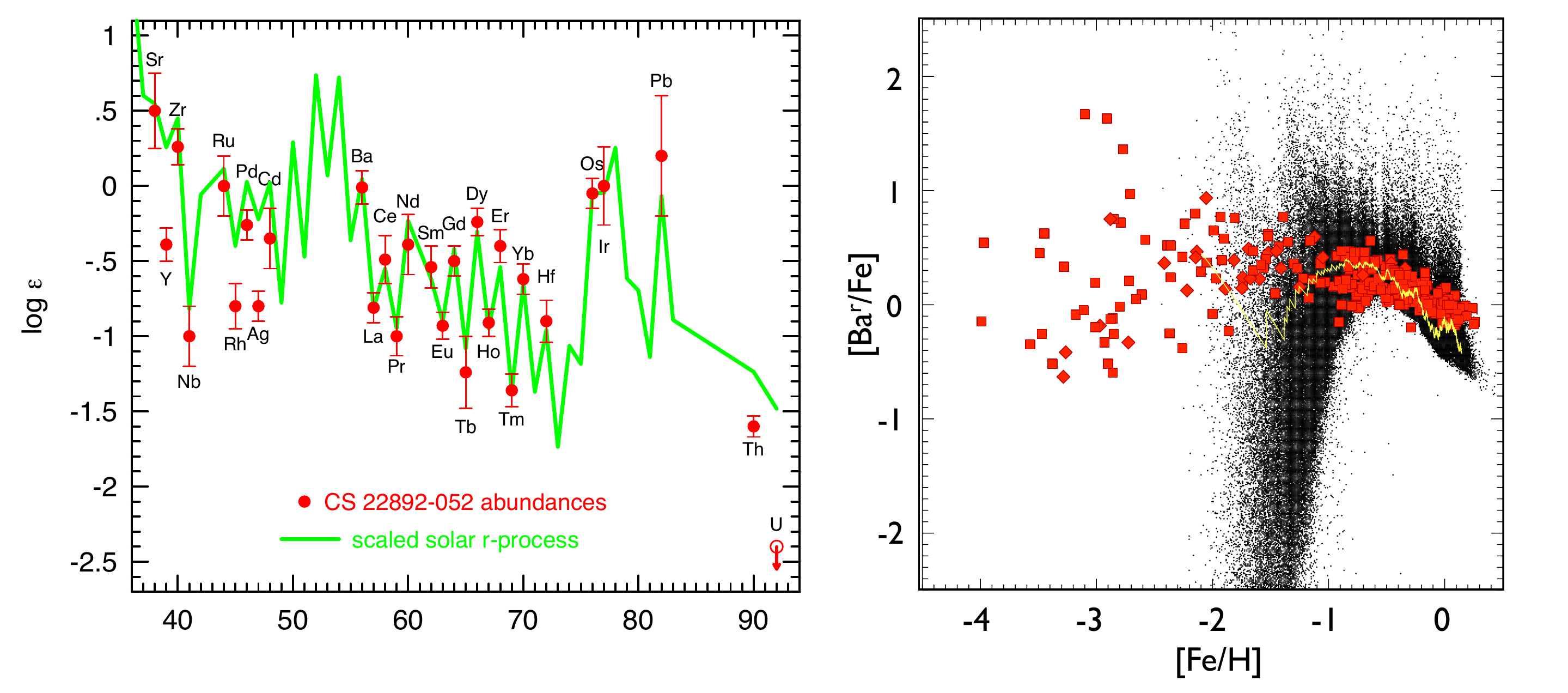}
\caption{Left: Metal abundances from metal-poor halo star CS 22892-052 compared to the scaled solar
distribution.  From \cite{Sneden}.  Right: The observed r-process fraction of Ba relative to Fe 
(red squares and diamonds) as a function
of [Fe/H], compared to the model calculations (black dots) of the evolution in the galactic rate of neutron star mergers,
assuming an average merger rate of 20/Myr and a coalescence timescale of 1 Myr.  From \cite{Argast}.}
\label{fig5:rprocess}
\end{center}
\end{figure} 

This presents quite a puzzle.  While there is an alternative and very attractive site that could be responsible
for the bulk of the r-process material found in the Milky Way -- neutron star mergers --
it is unlikely that this mechanism can account for the
r-process patterns we see in certain old, metal-poor halo stars (left panel, Fig. \ref{fig5:rprocess}).   Few
neutron-star mergers are expected at low metallicity because it takes time for the progenitor stars to
evolve, as the right panel of Fig. \ref{fig5:rprocess} shows.  Furthermore, because metal-poor stars sample
the r-process locally at a time when the galaxy is chemically inhomogeneous, one can estimate the rate of
r-process events in the early galaxy from the fluctuations in observed
abundances.  One finds $\sim$ $10^{-2}$/y.  While the uncertainty in this estimate is large, the rate is typical of core-collapse
supernovae and much higher than would be expected for neutron star mergers, even today.

Studies of metal-poor stars have been the main source of new information on the r-process.  The
metals on the surfaces of such stars are thought to have arisen from a single, or perhaps a few, nearby 
r-process sites.  The correspondence between the abundances in such stars and the scaled 
r-process distribution of our Sun (left panel, Fig. \ref{fig5:rprocess}) suggests that the r-process 
distribution produced in a single event is
similar to that averaged over many events.

Because of the difficulty of finding a robust r-process that naturally accounts for both the metal-poor-star data and
integrated r-process yields, my collaborators and I recently explored another scenario:
\begin{itemize}
\item  One r-process component that operates at early times, is associated with supernovae, and depends on the 
low-metallicity environments found in early massive stars.  This early r-process 
would need to account for the metal-poor-star r-process abundance pattern and the observed 
local enrichments.  It would need to operate until a metallicity of -2.5 to -2.0, the time a more robust 
r-process would then take over.
\item  The second process, unconstrained by the metal-poor star data but possibly responsible for
the bulk of r-process synthesis,  would then need to turn on in a manner consistent with r-process
chemical evolution studies. Neutron-star mergers would be the natural candidate for this second r-process.
\end{itemize}
There is also a candidate mechanism for the first r-process, neutrons produced by neutrino
reactions in the He shell of a type II supernova \cite{Epstein}.  This mechanism was originally suggested
as ``the r-process," studied in high-metallicity stars, and found to fail there because the 
neutron/seed ratio was about a factor of 20 too low \cite{WoosleyHaxton}.  This failure reflected both the 
relatively weak neutron source and the high abundance of neutron poisons such as $^{14}$N in the 
He shells of the progenitor stars that were studied.  The mechanism exploits the fact that neutrons
scatter off He without capturing, and
thus have an enhanced probability of capture on Fe and other heavy seeds. 

We re-examined this r-process in the context of metal-poor stars,
using a library of progenitors generated with the Kepler code,
and constructing more complete reaction networks.  The details are given in \cite{Banerjee}.  
Many of the conclusions of earlier studies were altered: the nuclear physics operated differently (for example,
neutral-current production of neutrons was found to be unimportant), and the effects of neutron poisons were
found to be less severe.  Most important, the study included the effects of neutrino oscillations.
With the assumption of an inverted hierarchy and a relatively hard heavy-flavor neutrino spectrum,
neutron abundances of $\sim 10^{19}$/cm$^3$were produced and maintained for long times in a variety of stars.  

\begin{figure}
\begin{center}
\includegraphics[width=16cm]{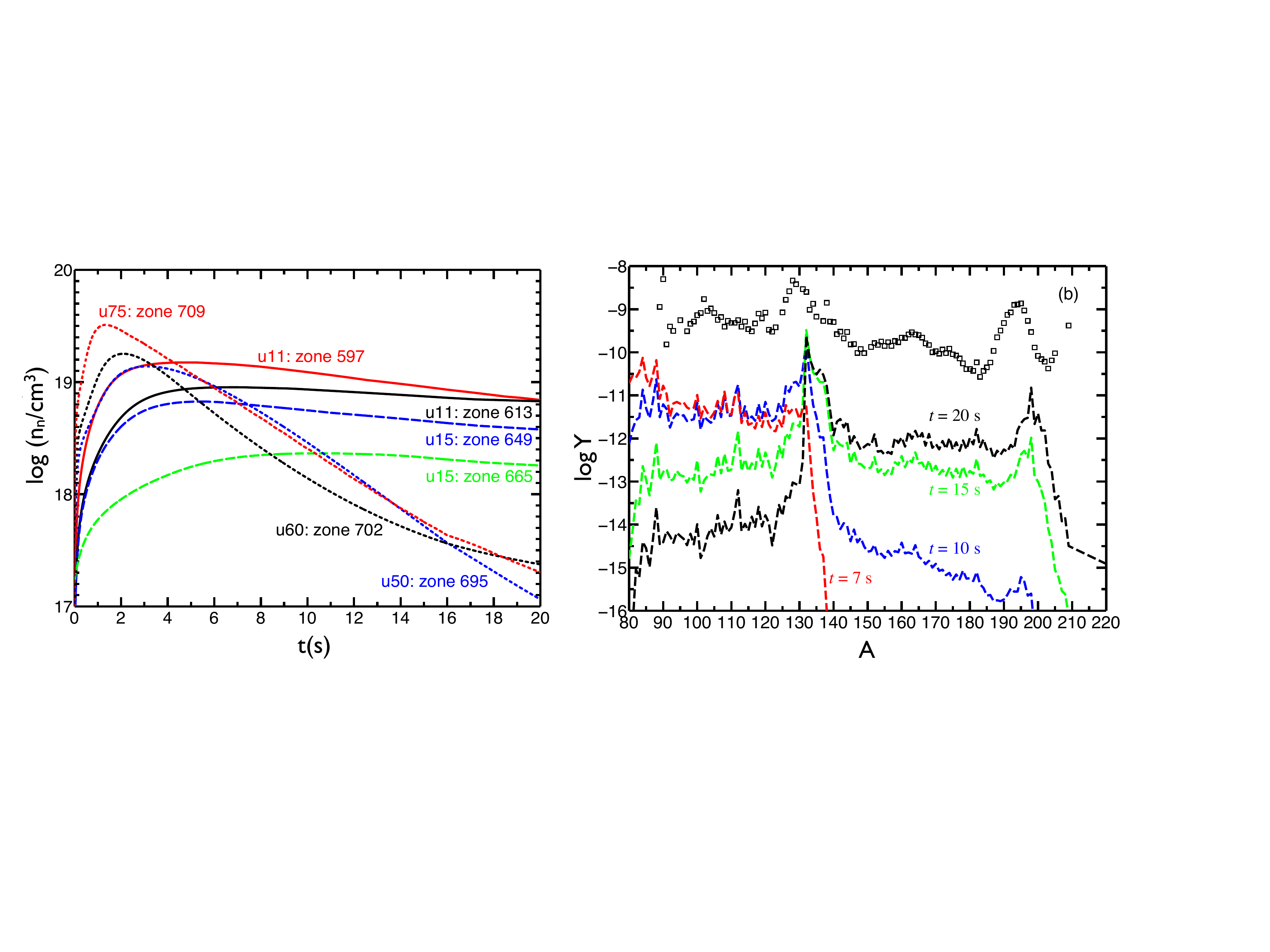}
\caption{Left: Neutron number densities as a function of time generated by neutrino interactions in the He
zone of allow-metallicity iron-core supernova.  The results are for specific zones in 
metal-poor progenitors between 11 and 75 M$_\odot$. Right: The resulting number fractions $Y_i(t)$ at
times of 7, 10, 15, and 20 s, compared to the solar r-process pattern (squares).}
\label{fig6:neutrino}
\end{center}
\end{figure} 

The mechanism is unusual:  It is a cold r-process, with
equilibrium established by neutron capture $\leftrightarrow$ $\beta$ decay, and with a path
farther from the neutron-drip line than in hot r-processes like the $\nu$-driven wind.   This has
consequences for the abundance pattern and for the time-development of the r-process (see Fig. \ref{fig6:neutrino}).  The time scales
are exceptionally long, with the third peak forming only after $\sim$ 20 seconds.  Furthermore, synthesis continues
after passage of the shock wave, augmented by the compression of the He shell and extending to times
$\sim$ 60 seconds (not shown in the figure).   The mechanism works
well for metallicities of $10^{-4} < Z < 10^{-3}$ in Fe-core supernovae between 11-16 M$_\odot$ and for
very massive progenitors of 49-75 M$_\odot$.  It naturally accounts for some of the features seen in metal-poor stars, 
such as a highly variable yield in Eu/Fe.  It is also interesting from
the neutrino physics viewpoint, turning on (inverted) or off (normal) with the choice of hierarchy.  I am not
aware of any other example in astrophysics as dramatically dependent on the neutrino hierarchy.

\section*{Acknowledgement}
I want to thank my key collaborators on the work reported here, including Aldo Serenelli, Liam Fitzpatrick,
Ami Katz, Projjwal Banerjee, Alex Heger, and Yong Qian, as well as the organizers for the opportunity to take
part in this celebration of the scientific career of a long-time friend, Jerry Draayer.
This work was supported in part by the U.S. Department of Energy
under contracts DE-SC00046548 (UC Berkeley) and
DE-AC02-98CH10886 (LBNL).

\end{document}